\documentclass[twocolumn,secnumarabic,amssymb, nobibnotes, aps, prd]{revtex4-2}

\usepackage{amssymb}
\usepackage{graphicx}
\usepackage{amsmath}

\usepackage{color}

\begin{document}

\title{Self-avoidant memory effects on enhanced diffusion\\ in a stochastic model of environmentally responsive swimming droplets}%

\author{Katherine Daftari}
\author{ Katherine A. Newhall}%
\email[corresponding author : ]{knewhall@email.unc.edu}
\affiliation{University of North Carolina at Chapel Hill}
\date{January 2021}%

\begin{abstract}

Enhanced diffusion is an emergent property of many experimental microswimmer systems that usually arises from a combination of ballistic motion with random reorientations.  
A subset of these systems, autophoretic droplet swimmers that move as a result of Marangoni stresses, have additionally been shown to respond to local, self-produced chemical gradients that can mediate self-avoidance or self-attraction.
Via this mechanism, we present a mathematical model constructed to encode experimentally observed self-avoidant memory and numerically study the effect of this particular memory on the enhanced diffusion of such swimming droplets.  
To disentangle the enhanced diffusion due to the random reorientations from the enhanced diffusion due to the self-avoidant memory, we compare to the widely-used active Brownian model. 
Paradoxically, we find that the enhanced diffusion is substantially suppressed by the self-avoidant memory relative to that predicted by only an equivalent reorientation persistence timescale in the active Brownian model.
We attribute this to transient self-caging that we propose is novel for self-avoidant systems.
Additionally, we further explore the model parameter space by computing emergent parameters that capture the velocity and reorientation persistence, thus finding a finite parameter domain in which enhanced diffusion is observable.
\end{abstract}

\maketitle

\section{Introduction}
Active particles are a class of nonliving nonequilibrium systems that derive their motility from
environmental energy consumption that is transformed into self-propulsion. 
There is great variety among motility-inducing forces, including chemical forces \cite{chemotaxis_synthetic_pioneer_1, chemotaxis_synthetic_pioneer_2, chemotaxis_synthetic_recent_1,chemotaxis_synthetic_recent_2}, photoelectric forces \cite{photoelectric_active_pioneer_1,photoelectric_active_pioneer_2,photoelectric_active_recent_1,photoelectric_active_recent_2}, and autophoresis \cite{autophoresis_synthetic_pioneer_1, Brujic2_autophoresis_synthetic_recent_1, Maass1_autophoresis_synthetic_recent_2, Brujic1_autophoresis_synthetic_recent_3, Maass2_autophoresis_synthetic_recent_4}.  
Such active systems are designed to mimic self-propulsion seen in microscale living systems, including the run-and-tumble behavior exhibited by bacteria \cite{run-and-tumble_pioneer_1,run-and-tumble_recent_1}, chemotactic responses to environmental stimuli \cite{chemotaxis_living_pioneer_1,chemotaxis_living_pioneer_2,chemotaxis_living_recent_1}, gravitactic responses \cite{gravitaxis_living_pioneer_1,gravitaxis_living_recent_1}, and photoelectric responses \cite{photoelectric_living_pioneer_1, photoelectric_living_pioneer_2,  photoelectric_living_recent_2}. 
For a comprehensive review of micro-scale active systems and current research developments, see Refs. \cite{Zhang, Ebbens, Ebbens_Howse}.

A hallmark feature of active systems is a ballistic movement, or ``swimming" that when interrupted by random and frequent directional changes gives rise to enhanced diffusion \cite{photoelectric_active_pioneer_2,photoelectric_active_recent_2,run-and-tumble_recent_1, Howse}.  
The biological advantage of enhanced diffusion is greater exploration of an area in a shorter period of time when compared to passive diffusion. 
 Consequently, biological effective diffusion and other single-particle emergent behaviors such as micro-scale transport, bacterial motion, and cell migration patterns and their biomimetic applications are research areas of great interest \cite{Zhang}.
 
In parallel, complete understanding of these phenomena via mathematical modeling provides design inspiration and permits cost-effective testing of novel systems; the most common model for active particles is the active Brownian particle (ABP) model. 
ABP combines directed motion 
resulting from a velocity dependent on the amount of available energy or ``fuel" with a rotational diffusion dependent on a defined persistence timescale, resulting in enhanced diffusion at time scales longer than the correlation time of the rotational diffusion \cite{Howse}.  
This model of competing ballistic and diffusive motion accurately predicts the enhanced diffusion of many experimental systems, 
such as those found in Refs.~\cite{chemotaxis_synthetic_pioneer_2, chemotaxis_synthetic_recent_1, chemotaxis_synthetic_recent_2, photoelectric_active_pioneer_2,photoelectric_active_recent_3,Ebbens_Howse,Howse}.

We are interested in the additional effect of spatio-temporal memory observed in slowly-dissolving autophoretic droplets \cite{Brujic2_autophoresis_synthetic_recent_1, Maass1_autophoresis_synthetic_recent_2, Brujic1_autophoresis_synthetic_recent_3, Maass2_autophoresis_synthetic_recent_4}, in addition to the persistence memory seen in ABP.
As these autophoretic droplets interact with the surfactant suspension, the particular physics induces a self-avoidant memory response.
Above a critical surfactant concentration, the leaking oily solute from the droplets is taken up into empty micelles.
This creates local heterogeneities in the surfactant concentration, which induce Marangoni stresses that cause the droplets to spontaneously swim in the direction of highest surfactant concentration.
This process continues as the droplets move, leaving behind a diffusing wake of solute-filled micelles and thereby a trail of depleted surfactant concentration.
It is precisely the fact that the diffusion of the micelles and surfactant is slow relative to the velocity of the droplets that causes self-avoidant motion as the droplets encounter gradients of solute concentration at the droplets' past locations that have not yet diffused away.  These past-history gradients induce Marangoni stresses that cause the droplets to move towards higher surfactant concentrations and therefore away from their past locations.

Despite being too large for the effects of thermal noise to be visible, the ballistic motion of the autophoretic experimental droplets is still punctuated by randomized directional changes, producing random-walk-like behavior. 
Such changes in direction reflect a transition between a dipolar (swimming) and a quadrupolar (stopped) hydrodynamic mode and the average frequency of these re-orientation events increases with P\'eclet number, droplet size, and the viscosity of the surrounding suspension \cite{Maass2_autophoresis_synthetic_recent_4}.
While this run-and-tumble-like behavior produces an enhanced diffusion that is consistent with the ABP model for the experimental parameters considered in~\cite{Brujic1_autophoresis_synthetic_recent_3}, we seek an understanding of the additional self-avoidant memory effect at play, particularly on the enhanced diffusion.

Motivated by the experimental system, we employ a model with a tunable memory response (which we distinguish from directional persistence) 
that qualitatively captures the essential features of the droplets and ignores the details of Marangoni stresses and hydrodynamic effects.
In this model, the particle is a mobile source of diffusing surfactant that descends its self-produced concentration gradient, resulting in a sustained ``swimming'' state and self-avoidant memory tied to the diffusion timescale.
To reproduce the coarse-grained effect of the random reorientations after each switch from the quadrupolar hydrodynamic mode of the experimental particles, we introduce thermal-like noise into the droplet's equation of motion.
This results in enhanced diffusion that intuitively one might expect the encoded self-avoidant memory to amplify as the particle evades its own past locations.  
However, we find the opposite: a suppression of enhanced diffusion over that predicted by an ABP with the same velocity and orientational persistence.  
We find evidence of {\em transient self-caging} as a possible explanation for this behavior.

In this paper, we begin in Sec.~\ref{sec:model} by presenting the mathematical details of this model for self-avoidant swimming droplets.
We investigate the memory effects of these model swimmers at long time scales by comparing the mean square displacement (MSD) to that of ABPs with the same velocity and orientational persistence in Sec.~\ref{III}.
To make these comparisons, we analytically derived an expression for the velocity in our model and numerically compute its orientational persistence timescale. 
We find that the equivalent ABP {\em overestimates} the enhanced diffusion of the model self-avoidant droplets, which we attribute to an unexpected side-effect of self-avoidant memory: transient self-caging.
In Sec.~\ref{IV} we further investigate the parameter space of the model, finding that with fixed noise strength, there is a limited regime of self-avoidant-memory strength within which enhanced diffusion is observable; the zero-memory limit of our model is not ABP.
We conclude the paper in Sec.~\ref{sec:conclusions}.

\section{A Model for self-avoidant Memory\label{sec:model}}

Motivated by the experimental system described previously, we propose a coupled model of a diffusion partial differential equation (PDE) for the surfactant concentration $c(\mathbf{x},t)$ and a stochastic differential equation (SDE) for the particle's location $\mathbf{X}(t)$.  
These equations are 
\begin{subequations}\label{eq:coupled}
\begin{equation} \label{eq:coupled PDE}
\begin{split}
\partial_t c(\mathbf{x},t) & = D \Delta c(\mathbf{x},t) +\alpha D R^2\delta_R(\mathbf{x}-\mathbf{X}(t)),
\end{split}
\end{equation}
for $\mathbf{x} \in \Omega \subseteq \mathbb{R}^2, t \geq 0$, and 
\begin{equation} \label{eq:coupled SDE}
\begin{split}
d\mathbf{X}(t) & = - \beta R \left(\int_{\Omega}\delta_R(\mathbf{x}- \mathbf{X}(t))\nabla_{\mathbf{x}} c (\mathbf{x},t)d \mathbf{x}\right) dt  \\
 & \quad \quad \quad \quad \quad \quad  + \sqrt{\sigma}d\mathbf{W}(t),
\end{split}
\end{equation}
\end{subequations}
with prescribed initial conditions ($c(\mathbf{x},0)=0$ and $\mathbf{X}(t) = \mathbf{0}$ unless otherwise noted) and boundary conditions (reflecting boundary conditions on $\partial \Omega$ unless otherwise noted).   

Equation \eqref{eq:coupled PDE} is a diffusion equation with diffusivity $D$ and a source term at the particle's current location. The time evolution of the concentration field holds the temporally-decaying memory of the particle's spatial history.
We note that the inclusion of diffusion on the source term differs from similar models for chemoatractive forces \cite{Lowen_2,Grima,Golestanian}. 
Motivation for this decision and the resulting effects are discussed later in this section.

In the source term with rate $\alpha D$, we introduce a ``size" $R$ to the particle using the radially-symmetric mollified delta function, $\delta_R(\mathbf{x}-\mathbf{X}(t))= \frac{1}{2\pi R^2}e^{-\frac{|\mathbf{x}-\mathbf{X}(t)|^2}{2R^2}}$.  (For the treatment of the particle as a point source with the Dirac delta function, see App.~\ref{App:dirac}; interestingly the particle does not swim.)
As the droplet releases oily solute from its membrane located at $|\mathbf{x}-\mathbf{X}(t)|=R$, this Gaussian emission pattern with standard deviation $R$ approximates the physical boundary of the particle while being more numerically and analytically tractable and does not require imposing a moving boundary condition on the concentration field to exclude the particle's interior.  
Since the particle's boundary does not physically exist in this model, Eq.~\eqref{eq:coupled PDE} also ignores the subtle effects the induced advection along the particle's surface has on the concentration gradient, as detailed in Ref.~\cite{Brujic2_autophoresis_synthetic_recent_1}.  We also keep $R$ fixed in time, thus we ignore depletion effects.

Equation \eqref{eq:coupled SDE} is a modified Langevin equation for a Brownian particle in a force field in the strong friction limit. 
We again mimic the size of the particle by convolving the gradient of the concentration field with the mollified delta function.  
We define the particle's response strength to the concentration gradient to be $\beta R$. 

As is conventional for an overdamped Langevin equation, $\mathbf{W}$ is a two-dimensional Weiner process scaled by the noise strength $\sqrt{\sigma}$. 
Recall, the experimental particles are athermal; this noise is to reproduce the stochasticity introduced by local fluctuations in the surfactant gradient that lead to re-orientations of the experimental droplet's swimming direction after each switch from a quadrupolar to bipolar mode.  
As the frequency of these re-orientation events depends on 
 P\'eclet number, droplet size, and the viscosity of the surrounding suspension \cite{Maass2_autophoresis_synthetic_recent_4}, the parameter $\sigma$ would likely be linked to other model parameters like diffusion $D$, droplet ``size'' $R$, and response to the concentration gradient $\beta$.  
 As we wish to keep noise effects constant to isolate the effects of self-avoidant memory in the present study, we ignore these possible dependencies in this study.

The stated model in Eq.~\eqref{eq:coupled} articulates the explicit relationship between the evolving concentration field and the particle trajectories. 
As the particle moves, its emissions induce changes in the local concentration field and it leaves behind diffusing physical evidence of its trajectory. 
Thus, the historical information or memory of the particle’s past locations is contained within the current state of the evolving concentration field.
The memory encoded in the concentration field allows each particle to “remember” where it has been (hotspots in the concentration field) and avoid its past trajectory with response strength decreasing as the time lag increases. 
Via integration of the whole gradient field at each time point, the particle becomes spatially omniscient as it moves with dependence on the affecting forces from every spatial point on the domain. 
In time, the particles are pseudo-omniscient as their ability to ``see" into the past through interaction with the concentration gradient diminishes exponentially in time. 
This unique behavior abolishes time-reversal symmetry although the coupled configuration in Eq.\eqref{eq:coupled} is Markovian since there is no explicit dependence on the trajectory’s past steps.

To limit the number of parameters under investigation, we nondimensionalize Eq.~\eqref{eq:coupled}. 
We choose $R$ as a natural length scale and non-dimensionalize $c$ without any scaling for simplicity. 
Temporarily, we leave time scale $T$ arbitrary.
Under the scalings $\mathbf{y}=\frac{\mathbf{x}}{R}$, $\mathbf{Y}=\frac{\mathbf{X}}{R}$, $\tau= \frac{t}{T}$, and $\mathbf{B}=\frac{\mathbf{W}}{\sqrt{T}}$, we arrive at
\begin{subequations} \label{eq:nondim coupled}
\begin{equation} \label{eq:nondim coupled PDE}
\begin{split}
\partial_t c(\mathbf{y},t) & = \mu\Delta c(\mathbf{y},t) +\mu \phi \exp \left[ -\frac{|\mathbf{y}-\mathbf{Y}(t)|^2}{2}\right], 
\end{split}
\end{equation}
for $\mathbf{y} \in \Omega, \quad t \geq 0$, and
\begin{equation} \label{eq:nondim coupled SDE}
\begin{split}
d\mathbf{Y}(t) & = - \nu \left(\int_{\Omega}\exp \left[ -\frac{|\mathbf{y}-\mathbf{Y}(t)|^2}{2}\right] \nabla_{\mathbf{y}} c (\mathbf{y},t)d \mathbf{y}\right)dt  \\
 & \quad \quad \quad \quad \quad \quad  +\sqrt{\epsilon} d\mathbf{B}(t)
\end{split}
\end{equation}
\end{subequations}

\noindent
where $c$, $t$ and $\Omega$ are re-used for their non-dimensional versions for convenience. 
We have mapped the dimensional parameters as follows: $D\rightarrow \mu=\frac{DT}{R^2}$, $\alpha \rightarrow \phi = \frac{\alpha R^2}{2\pi}$, $\beta \rightarrow \nu =\frac{\beta T }{2\pi  R}$,
and $\sigma \rightarrow \epsilon=\frac{\sigma T}{R^2}$. 

We note that a typical time scale for the diffusion equation is $T=\frac{L^2}{D}$. 
Although traditional, this choice would prevent us from seeing directly the effects of changing $D$, which encodes the memory timescale. 
Increasing diffusivity would contract time such that the past-history wake of the particle would adjust to decay at the same rate.  
Thus, to observe the effects of this memory-encoding diffusivity,  we choose to fix the  stochastic diffusivity $\epsilon =0.75^2$, thereby  choosing $T=0.75^2  \frac{R^2}{\sigma}$. 
Keeping the value of $\sqrt{\epsilon}$ fixed at $\frac{3}{4}$ was a convenient choice made to maintain balance between the stochastic effects, controlled by $\sqrt{\epsilon}$, and the deterministic effects of swimming as well as self-avoidant memory, controlled by $\nu$ and $\mu$.

We can simplify the system by taking the Fourier transform and solving the PDE \eqref{eq:nondim coupled PDE} on an infinite domain, $\Omega = \mathbb{R}^2$, explicitly. 
Incorporating this solution into the SDE \eqref{eq:nondim coupled SDE} we arrive at 
 the mathematically equivalent system for the particle in an infinite domain 
\begin{equation} \label{eq:nondim combined}
\begin{split}
 d\mathbf{Y} & =   \frac{\pi}{2}\mu \nu \phi \int_{0}^{t}\exp \left[ -\frac{|\mathbf{Y}(t)-\mathbf{Y}(s)|^2}{4(1+ \mu(t-s))}\right] \\
 & \cdot (1+ \mu(t-s))^{-2} (\mathbf{Y}(t)-\mathbf{Y}(s))ds  dt + \sqrt{\epsilon} d\mathbf{B};
\end{split}
\end{equation}
see App.~\ref{App:Solve} for details.
This non-Markovian SDE explicitly reveals the dependence on all the particle's previous locations
 via integration in time; 
the exponential kernel decays in both time and space, revealing $\mu^{-1}$ to be the self-avoidant memory timescale.  In this way the model contains a self-avoidant memory, one of the key features of the experimental droplets, with controllable timescale $\mu^{-1}$.  In addition to producing a self-sustained swimming state, the form of this force also allows the droplets to hover above a bottom plate with the addition of gravity to the model (see App.~\ref{App:hover} for details) in much the same way that the experimental droplets do~\cite{Brujic2_autophoresis_synthetic_recent_1}.

The formulation of the model in Eq.~\eqref{eq:nondim combined} is convenient for simulation since it does not require solving the PDE on a large domain to capture long-time dynamics.  It additionally removes the integral in the SDE over $\mathbb{R}^2$ and replaced it with an integral over $t$.
We integrate Eq.~\eqref{eq:nondim combined} in time with the Euler-Maruyama method while using Simpson's rule to integrate the memory kernel at each step. 
This algorithm is a first order method.

The limiting behavior of these two equivalent systems, Eq.~\eqref{eq:nondim coupled} and~\eqref{eq:nondim combined}, foreshadows their distinction from the active Brownian model since it reveals that removing the distinguishing feature of memory by taking $D\to\infty$ will not reduce our model to ABP.  
The parameter $D$ was added to the source term in Eq.~\eqref{eq:coupled PDE} to achieve balance between the rate at which the oil diffuses and the rate at which the oil is expelled in this limit.  (If instead the source term remained constant relative to $D$, then it would effectively vanish in the limit of $D\to\infty$.)
The nondimensional parameter $\mu$ therefore appears on the source term in Eq.~\eqref{eq:nondim coupled PDE}, and
$\partial_t c(\mathbf{y},t)\to\infty$ as $\mu\to\infty$.  To leading order, the concentration field satisfies 
 the 
Poisson equation
\begin{equation}\label{eq:Poisson}
   \Delta c(\mathbf{y})= -\phi \exp \left[ -\frac{|\mathbf{y}-\mathbf{Y}(t)|^2}{2}\right]  \qquad \mu\to\infty.
\end{equation}

The concentration field is now memory-less since it instantaneously equilibrates as the particle moves.
On an infinite domain, the solution to Eq.~\eqref{eq:Poisson} will be radially symmetric around the particle's location, and therefore the integral in Eq.~\eqref{eq:nondim coupled SDE} will always be zero.
As a result, particles  experience motility solely from thermal fluctuations, namely simple Brownian motion.

Also noteworthy is the ``full memory'' limit of Eq.~\eqref{eq:nondim coupled PDE} which is $\partial_t c \to 0$ as $\mu\to 0$. 
The concentration field remains fixed at its initial conditions as the source term and the diffusion term vanish in this limit. 
The particle experiences thermal fluctuations while responding to the concentration gradient of the fixed concentration landscape, thereby statistically preferring concentration minima. 
The steady state (if one exists) would be almost solely determined by the initial topography of $c(\mathbf{y})$ and the relative size of $\epsilon$. 
In fact, the entire coupled system in Eq.~\eqref{eq:nondim coupled} reduces to simple memoryless Brownian motion 
\begin{equation}\label{eq:mu0}
 d\mathbf{Y}  = \sqrt{\epsilon} d\mathbf{B} \qquad \mu\to 0
\end{equation}
in this limit, as the source term which encodes the memory vanishes.
This is consistent with Eq.~\eqref{eq:nondim combined} which also reduces to simple Brownian motion in the same limit $\mu\to 0$ when it is assumed that the initial concentration field is constant. 
Therefore we focus our study on intermediate range of $\mu$ where the effects of the noise, the swimming, and the memory are all observable.
The limiting behaviors of our model as described above can all be traced back to the addition of a second $\mu$ on the source term. 
As stated previously, inclusion of a diffusive scaling on the source term was required to ensure that the source term remained in the limit as $D \rightarrow \infty$, in the hope that the model would revert to active Brownian motion with no self-avoidant memory. 
In our results, we discuss the role of this diffusive scaling in generating previously unseen behaviors in this class of models.

\section{Comparative Analysis\label{III}}

\begin{figure}
    \includegraphics[width = 0.78\textwidth]{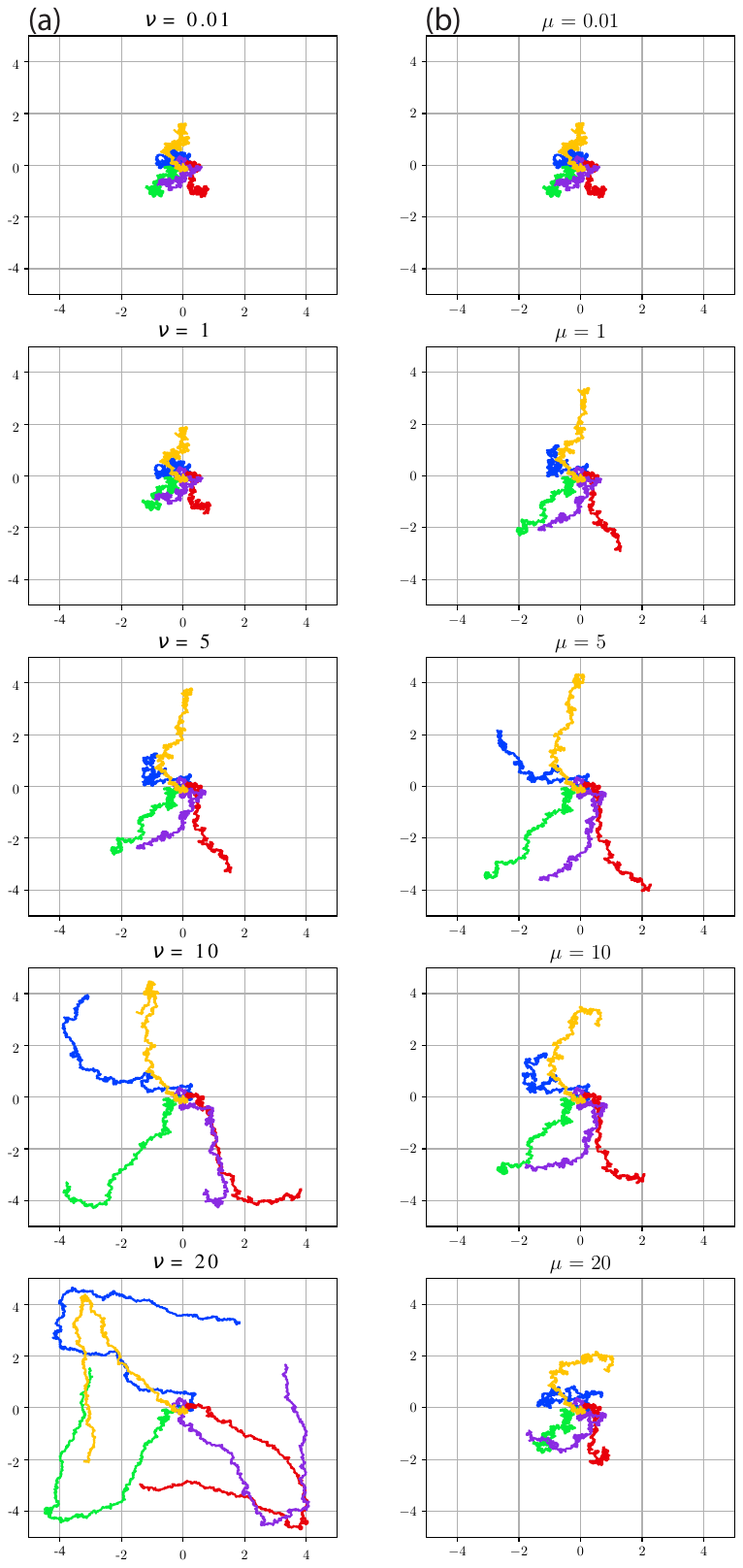}
    \caption{Numerical simulations of Eq.~\eqref{eq:nondim coupled} were carried out using a forward-time centered-space finite difference scheme for the PDE and the Euler-Maruyama method for integrating the SDE where the trapezoid rule was used to compute the integral therein. We confine $\mathbf{X}(t)$ to a box $B= \{x \in (-5,5), y \in (-5,5)\}$ with insulating boundary conditions such that $\frac{\partial c (\mathbf{x},t)}{\partial t}=0 \quad \forall \mathbf{x} \in \{\partial B\}$. This has the effect of reflecting the particle back into the domain when it reaches the boundary. The initial condition is $c(\mathbf{x},0)=0$. Note that these trajectories are visually indistinguishable from active Brownian motion. (a) For $\phi = 1$, $\mu = 5$ and a noise level of $\sqrt{\epsilon} = 0.75$, we see the dominant effect of $\nu$ which is to increase the velocity. (b) For $\phi = 1$, $\nu = 7$, and a noise level of $\sqrt{\epsilon} = 0.75$, we see the effects of $\mu$ which primarily increases the turning frequency and has a secondary effect on velocity.}
       \label{fig:paths}
  \end{figure}

Numerically simulated trajectories of the coupled model given by Eq.~\eqref{eq:nondim coupled}  are shown in Fig.~\ref{fig:paths}. 
These trajectories illustrate the main features of active matter: a swimming velocity with a slowly diffusing direction. 
Increasing $\nu$, and therefore the response to the concentration gradient causes the particle to swim faster, shown in Fig.~\ref{fig:paths}a, while increasing $\mu$, and therefore shortening the timescale of the diffusion (decreasing the memory), has a secondary effect on the velocity, but also causes the particles to turn faster, shown in Fig.~\ref{fig:paths}b. 
An increase in turning frequency was also observed experimentally in \cite{Brujic1_autophoresis_synthetic_recent_3} as surfactant concentration was increased, prompting a transition from ballistic motion to diffusion. (See Figure 2 in \cite{Brujic1_autophoresis_synthetic_recent_3}. Recall the Marangoni effect which causes the droplets to ``search" for areas of higher surfactant concentrations, while the droplets simultaneously modify the local concentration. )

We seek to look beyond the combined effects of swimming and random directional changes in producing enhanced diffusion and understand the additional effects of self-avoidant memory. 
Specifically, we compare our model to ABP given by the equations
\begin{subequations}\label{eq:ABP}
    \begin{equation}
               dX = V\cos(\theta(t))dt + \sqrt{\epsilon}dW_x
   \end{equation}
   \begin{equation}
                dY = V\sin(\theta(t))dt + \sqrt{\epsilon}dW_y
    \end{equation}
    \begin{equation}
        d\theta = \frac{1}{\sqrt{\tau}}dW_{\theta}
    \end{equation}
\end{subequations}
where $\sqrt{\epsilon}$ is the strength of the additive noise in each spatial component (consistent with the model in Eq. \eqref{eq:nondim combined}), $V$ is the swimming velocity, and $\tau$ is the persistence timescale of the rotational diffusion \cite{Brujic1_autophoresis_synthetic_recent_3, Howse, Lowen}. 
These latter two parameters do not explicitly appear in our model; we will compute them and compare the MSD of the two models to understand the effects of self-avoidant memory on enhanced diffusion.  

In Sec.~\ref{sec:velocity} we present an analytic equation that is numerically solved for the velocity of the swimming solution to Eq.~\eqref{eq:nondim combined}.  
This velocity is consistent with the intermediate ballistic regime of the MSD, computed numerically for Eq.~\eqref{eq:nondim combined} and given by
\begin{equation}\label{eq:MSD ABP}
    \mathbb{E}[\mathbf{X}(t)^2]=4V^2 \tau^2\left[2 \left(e^{-\frac{ t}{2\tau}}-1\right)+ \frac{ t}{\tau} \right] + 2 \epsilon t
\end{equation}
for Eq.~\eqref{eq:ABP}.  
In Sec.~\ref{sec:tau} we determine $\tau$ by numerically computing the orientation correlation function but find that the memory induced from modifying the environment causes a reduced effective diffusion as compared to ABP with identical angular persistence.

\subsection{\textbf{Intermediate Time Scales: Ballistic Motion}\label{sec:velocity}}

Unlike active Brownian models, the proposed model has a non-explicit intrinsic velocity; directed motion at this velocity may become observable at intermediate time scales under appropriate conditions for $\nu$ and $\mu$. 
To find an analytic form for the velocity, we seek a deterministic constant velocity (``steady state") solution to the combined model Eq.~\eqref{eq:nondim combined}.
Without loss of generality, suppose  $\mathbf{Y}(t)=\langle Vt,0 \rangle $; thus $V$ must solve 
\begin{equation}\label{eq:V int first}
\begin{split}
    \frac{d\mathbf{Y}}{dt}= V = &  \frac{\pi}{2}\mu \nu \phi \int_{-\infty}^{t}\exp \left[ -\frac{|Vt-Vs|^2}{4(1+ \mu(t-s))}\right] \\
    & \cdot (1+ \mu(t-s))^{-2} (Vt-Vs)ds .
\end{split}
\end{equation}
\noindent
Under the transformation $z=\mu(t-s)$, the constant velocity $V$ therefore satisfies

\begin{equation}\label{eq:V int final}
    1= \frac{\pi}{2}\frac{\nu}{\mu}\phi \int_{0}^{\infty} \frac{z}{(1+z)^{2}} \exp \left[- \left(\frac{V}{\mu} \right)^2 \frac{z^2}{4(1+z)} \right]dz.
\end{equation}

\begin{figure*}
     \centering
      \includegraphics[clip, trim=0.5cm 4.5cm 0.5cm 4.5cm, width=0.95\textwidth]{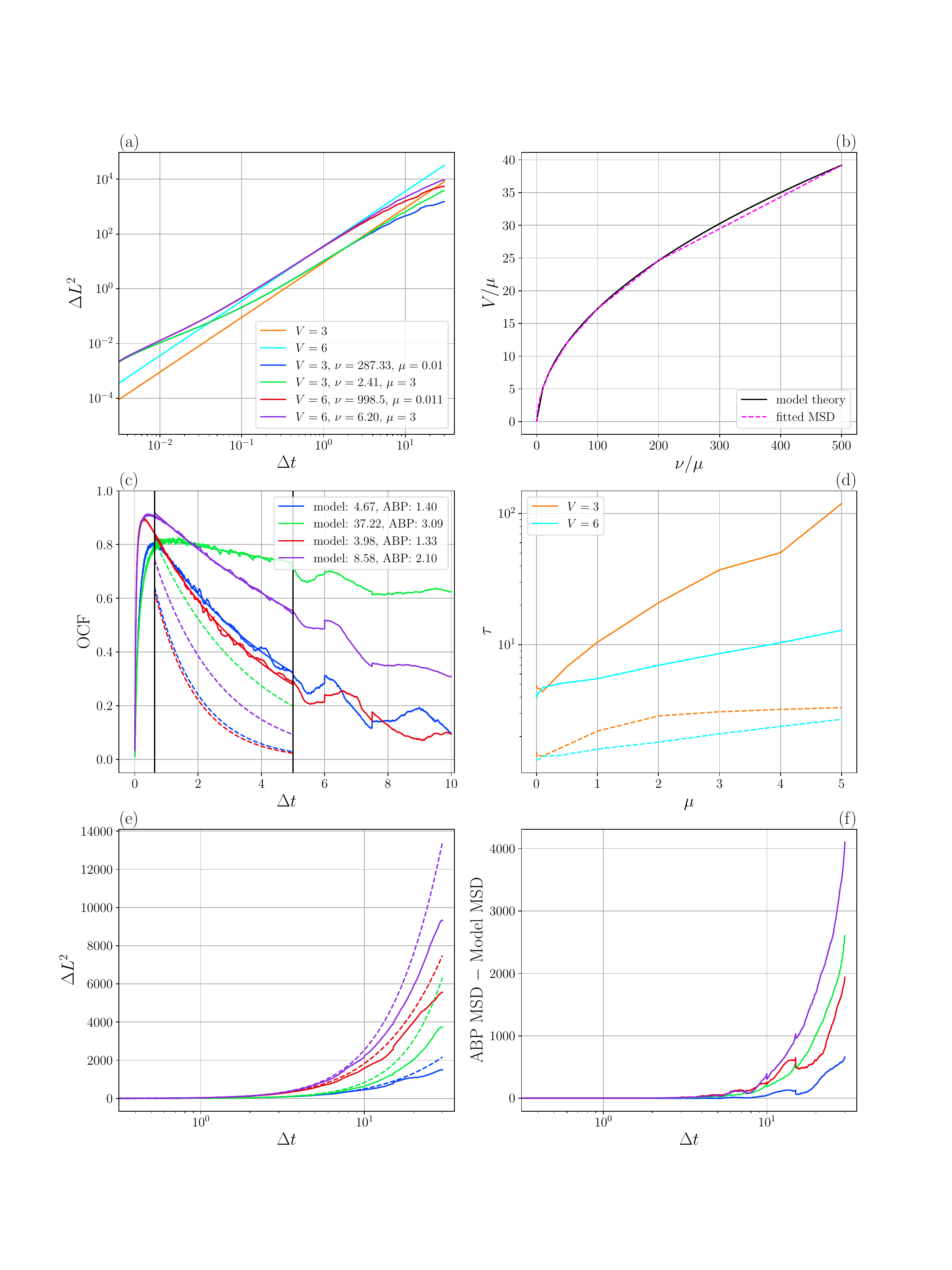}
     \caption{(color online) (a) Empirically computed MSDs of the model in comparison to benchmark pure ballistic motion. Three distinct regions of the MSD are visible: classical diffusion, directed motion in alignment with the benchmark ballistic lines, and enhanced diffusion where the MSD pulls away from the ballistic motion. Legend colors are consistent with other panels throughout the rest of the figure. Legend colors correspond to all panels. 
     (b) Collapsed model velocity theory from solving Eq.~\eqref{eq:V int final} in comparison to fitted values extracted from the ballistic regime of the MSD of the form $V^2 t^2$. (c) Model OCFs (noisy solid lines) with fitted $\tau$ (solid smooth curves) with comparison to $\tau$ fitted from the enhanced diffusion regime of the MSD consistent with ABP Eq.~\eqref{eq:ABP MSD big t} (dashed smooth curves). Values of $\tau$ for each corresponding color are given in the legend. (d) Both fitted $\tau$ values from ABP MSD (dashed) and OCF of our model (solid)  as they vary over $\mu$. (e) Model MSD (solid) and ABP MSD Eq.~\ref{eq:MSD ABP} (dashed) under the same $\tau$ fitted to the model OCF and theoretical $V$. (f) Signed difference between model and ABP MSDs in panel (e). }
     \label{fig:Big Fig}
 \end{figure*}

\noindent
For each value of $\frac{\nu}{\mu}$, we solve for the value of $\frac{V}{\mu}$ that makes the above integral equal to 1
numerically in Python with \verb|scipy|.  Under the change of variables $x=\frac{2}{\pi} \arctan(z)$, we map the domain $(0,\infty)$ to $(0,1)$ for ease of numerical integration.
The resulting monotonically increasing dependence of $\frac{V}{\mu}$ on $\frac{\nu}{\mu}$ is plotted as the solid black line in Fig.~\ref{fig:Big Fig}b.
Alternately, we can select $\mu$ and $V$ and compute $\nu$ to satisfy Eq.~\eqref{eq:V int final}.

We can directly compare this theory to ABP on the timescale at which ballistic motion is dominant. 
It is evident from Fig.~\ref{fig:Big Fig}a that the ballistic portion of the simulated MSD aligns with the computed velocity from Eq.~\eqref{eq:V int final}.
At such small times, the MSD of ABP asymptotically reduces to
\begin{equation}\label{eq:MSD ABP small t}
    \mathbb{E}[\mathbf{X}(t)^2] \sim V^2t^2 + 2\epsilon t 
\end{equation}

\noindent
as $t \rightarrow 0$ (see App.~\ref{App:smallMSD} for details.) 
Fitting $V$ from the ballistic portion of the MSD of our particles is also in good agreement with the theory, as shown in Fig.~\ref{fig:Big Fig}b. 

\begin{figure*}

    \includegraphics[width = 0.95\textwidth,clip, trim=0 8.5cm 0 0]{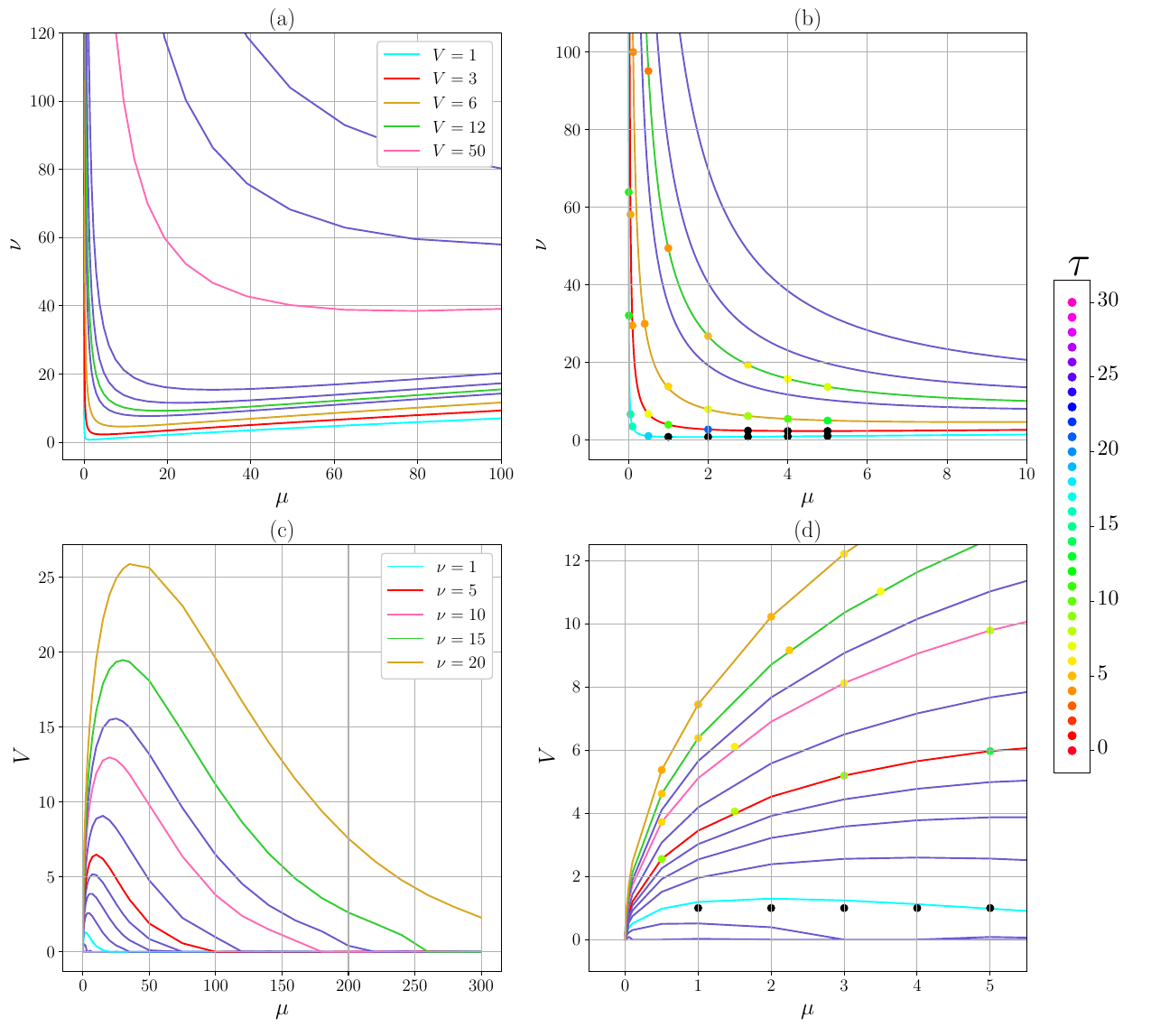}

    \caption{(color online) Visualizations of the four dimensional parameter space of $V$ from solving Eq.~\eqref{eq:V int final}, $\tau$ from fitting the model OCF, along with model parameters $\nu$ and $\mu$. (a) and (b) depict constant velocity $V$ contours, while in (c) and (d) constant $\nu$ contours are shown.  Panels (b) and (d) show sub-regions of panels (a) and (c), respectively, with added individual points depicting values of $\tau$ given by the color bar. Note that $\tau$ values 
    larger than the experiment length (greater than 30) are colored black.}
    \label{fig:contour fig}
\end{figure*}

We point out that the existence of an observable ballistic regime in the MSDs from our model requires a sufficient swimming velocity $V$ to dominate the additive noise.  
In the ABP model, this can be guaranteed by changing the stated parameter $V$, whereas in our model, there must be consideration for the parameters $\mu$ and $\nu$ due to the explicit functional relationship $V=f(\mu, \nu)$ given by Eq.~\eqref{eq:V int final}.
To see this functional relationship more clearly, the contours of constant velocity are plotted in Fig.~\ref{fig:contour fig}a and the contours of constant $\nu$ are plotted in Fig.~\ref{fig:contour fig}c.  
These figures agree with the limits from Sec.~\ref{sec:model} in that $V=f(\mu, \nu) \rightarrow 0$ 
when taking either $\mu\to0$ or $\mu\to\infty$ with fixed $\nu$, and the model system Eq.~\eqref{eq:nondim coupled} or Eq.~\eqref{eq:nondim combined} reduces the particle motion to simple Brownian motion.  
Furthermore, taking $V\to 0$ in Eq.~\eqref{eq:V int first} results in a divergent integral; for the integral to converge, either $\mu$ or $\nu$ in the prefactor must also go to zero. 
The result is no transition to swimming at a small finite value of these parameters.  
Similarly, the integral also diverges as $\mu\to \infty$. Figure~\ref{fig:contour fig}c most clearly shows the relevant intermediate values of $\mu$ for which a significant velocity exists and ABP-like motion with a ballistic regime is expected for the model system. 

Figure~\ref{fig:contour fig}a most clearly shows that 
for any given velocity, there exists a minimal $(\mu, \nu)$ pair. 
If we interpret this in the context of the coupled model given by Eq.~\eqref{eq:nondim coupled}, it suggests the existence of an optimal response strength and diffusivity pairing which act on the particle to produce directed motion at a specific speed. 
Moving off of this minimum illustrates the parameter couplings which must balance to keep the particle moving at a given speed. 
For example, decreasing memory (increasing $\mu$) allows the particle's trail to diffuse faster which weakens local gradients, and thus requires that the response strength to the weakened gradient be increased (increasing $\nu$).

\subsection{\textbf{Long Time Scales: Enhanced Diffusion}\label{sec:tau}}

Figure \ref{fig:Big Fig}a shows a departure of the MSD from ballistic motion at longer time scales.  For ABP, this departure happens at time scales $t\gg \tau$ for which the MSD \eqref{eq:MSD ABP} is asymptotic to 
\begin{equation}\label{eq:ABP MSD big t}
    \mathbb{E}[\mathbf{X}(t)^2] \sim (4V^2\tau  + 2\epsilon) t
\end{equation}
as $t \rightarrow \infty$.  
Particle reorientations that decorrelate with timescale $\tau$ enhance the diffusion term $2\epsilon t $ with the term $4V^2\tau$.  

To estimate $\tau$ from trajectories given by our model
we  first numerically compute the normalized orientation correlation function (OCF) which measures the relative angle between consecutive movements. It is given by
\begin{equation}\label{eq:OCF}
    C(\Delta t)= \left< \frac{\mathbf{v}(t)\cdot \mathbf{v}(t+ \Delta t)}{|\mathbf{v}(t)||\mathbf{v}(t+ \Delta t)|} \right>_{t}
\end{equation}
where $\mathbf{v}(t)= \mathbf{Y}(t)- \mathbf{Y}(t- \Delta t)$ is the directional displacement between times $t$ and  $t-\Delta t$ (see App.~\ref{app:OCF} for details).  This function computed for the trajectories is shown in Fig.~\ref{fig:Big Fig}c as the noisy solid lines.  
Note that $C(\Delta t)\to 0$ as $\Delta t \to 0$ because the motion at such small timescales is dominated by the uncorrelated additive noise.  As $\Delta t$ increases ballistic motion starts to dominate which is reflected in the OCF that approaches values near unity.   
The portion of $C(\Delta t)$ displaying exponential decay, due to the transition to enhanced diffusion at even longer $\Delta t$, is fit by a single exponential given by $C(\Delta t)= e^{-\frac{\Delta t}{\tau}}$ as is consistent with ABP \cite{ Brujic1_autophoresis_synthetic_recent_3, Lowen}. 
These fits are shown by the smooth solid lines in Fig.~\ref{fig:Big Fig}c and the resulting values of $\tau$ as a function of $\mu$ in Fig.~\ref{fig:Big Fig}d.

While the model OCF is well fit by an exponential decay, the long time asymptotics of the MSD given by Eq.~\ref{eq:ABP MSD big t} in Fig.~\ref{fig:Big Fig}e (dashed lines) shows that ABP substantially overestimates the enhanced diffusion of our model (solid lines).  
This overestimation is larger for the two larger values of $\mu$ that correspond to weaker self-avoidant memory, as shown in Fig.~\ref{fig:Big Fig}f.  
Alternatively, using $V$ from Eq.~\eqref{eq:V int final}, $\tau$ is determined by fitting the long time MSD to Eq.~\eqref{eq:ABP MSD big t}.  
These values of $\tau$, plotted as the dashed lines in Fig.~\ref{fig:Big Fig}d, substantially underestimate the decorrelation time scale of our model, also shown by the corresponding dashed lines of exponential decay in Fig.~\ref{fig:Big Fig}c.
Although the form of exponential decay of the orientational persistence is consistent with ABP and quantifiable by $\tau$, it alone is not enough to predict the enhanced diffusion of our model. 
There are additional effects of self-avoidant memory beyond the persistence memory, which is the only memory present in ABP. 

At constant velocity, the effect of increasing self-avoidant memory (decreasing $\mu$) is seen in Fig.~\ref{fig:Big Fig}a.  
The MSDs with smaller $\mu$ in both cases depart from the ballistic regime earlier, and thus exhibit less enhanced diffusion.  
This corresponds to Fig.~\ref{fig:Big Fig}d where for smaller $\mu$ the OCF decays more rapidly as measured by a smaller value of $\tau$.
This is further illustrated in Fig.~\ref{fig:contour fig}b, showing a decrease of $\tau$ with decreasing $\mu$ along contours of constant velocity. 
For fixed $\mu$, $\tau$ increases with decreasing $\nu$ (although velocity decreases).
Thus we see that one effect of self-avoidant memory as it is present in our model is to decrease orientational persistence: swimmers with high memory experience weak orientational persistence and vice versa. 

\begin{figure}
\vspace{-1cm}
    \centering
    \includegraphics[scale=0.4]{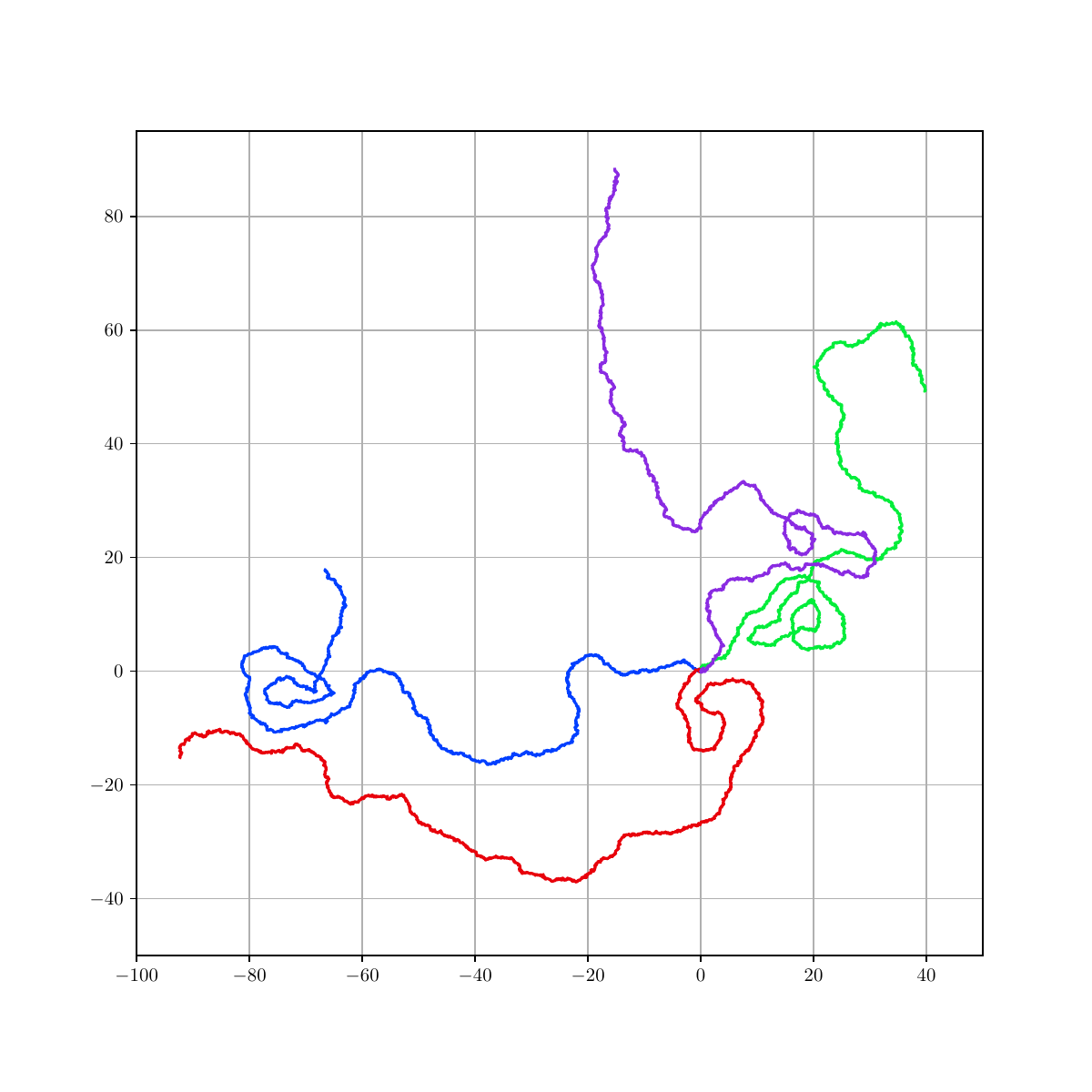}
    \vspace{-0.75cm}
    \caption{With $\mu = 0.01$ and $V = 6$, four sample paths are shown which illustrate the caging of enhanced diffusion experienced due to high memory.}
    \label{fig:twisty paths}
\end{figure}

A more exotic effect of self-avoidant memory is shown by the trajectories in Fig.~\ref{fig:twisty paths} and provides a plausible explanation for the surprising fact that ABP overestimates the enhanced diffusion of the model.
To avoid crossing their own self history, paths turn back on themselves and continue turning inward, becoming caged for a while before 
enough diffusion has occurred for them to leave this self-created trap.  
This transient self-trapping perhaps explains the reduced enhanced diffusion as compared to ABP with equivalent orientation persistence time. 
Self-trapping has been studied in autophoretic systems like that which we model, although it has only been found in chemoattractant experimental systems \cite{Liebchen_and_Lowen, Tsori} and model systems \cite{Lowen_2, Grima, Golestanian} with self-attracting memory. 
It will be interesting to find out whether self-avoidant experimental systems like that in \cite{Brujic2_autophoresis_synthetic_recent_1} show similar self-trapping.  

 \section{Limiting Behavior\label{IV}}

Since the relevant experimental systems are well described by ABP and we can explicitly tune memory in our model, we anticipated that we could find a parameter regime (low memory, high $\mu$) in which the enhanced diffusion produced by our model was well described by ABP.  
However, as discussed at the end of Sec.~\ref{sec:model}, the limiting behavior of systems \eqref{eq:nondim coupled} and \eqref{eq:nondim combined} as $\mu\to\infty$ is simple Brownian motion, indicating that enhanced diffusion with low self-avoidant memory may not be possible.  
We revisit this limit with further simulations in light of the emergent parameters $V$ and $\tau$, considering both $\nu\to\infty$ with $V$ fixed as well as dynamic $V$ with $\nu$ fixed.  
Additionally, we investigate the high memory limit ($\mu\to 0$), and find it consistent with Eq.~\eqref{eq:mu0} describing classical Brownian motion with (unfortunately) no further memory effects to investigate.

From Figs.~\ref{fig:contour fig}a and b, we can consider the limit most likely to produce enhanced diffusion consistent with ABP: removing the memory via the limit $\mu \rightarrow \infty$  while keeping the particle at constant velocity by fixing $V$.
Visually we observe that as $\mu \rightarrow \infty$, the contours of $V$ become flatter, reproducing the behavior seen in Fig.~\ref{fig:paths} which shows that ballistic motion is sensitive to changes in the gradient response $\nu$.  
Moreover, remaining on one $V$ contour requires $\nu\to\infty$ much slower than $\mu\to\infty$.
To investigate the enhanced diffusion in this limit, we look at $\tau$ in Fig.~\ref{fig:contour fig}b.  
Following a $V$ contour as $\mu\to\infty$ results in an increase in $\tau$ corresponding to longer orientational persistence (or less change in direction).
 
As a result of decreasing the self-avoidant memory timescale ($\mu\to\infty$ while maintaining a constant velocity $V$), we find that both the past history of the trajectory and the Brownian noise become less important in influencing the future location of the trajectory.
Furthermore, with the addition of an increased gradient response by taking $\nu\to\infty$ (as required to keep the particle at constant $V$), the deterministic gradient response force in Eq.~\eqref{eq:nondim coupled} dominates the Brownian noise, and consecutive steps become more correlated.   
This increases the persistence time $\tau$, and the trajectories approach purely ballistic motion with no enhanced diffusion at observable finite times.  

Figs.~\ref{fig:contour fig}a and b also allow for considering infinite memory ($\mu \rightarrow 0$), again with constant velocity.
In Fig.~\ref{fig:contour fig}a, we see that the gradient response required (given by the size of $\nu$) to keep the particle moving at constant velocity $V$ rapidly blows up to $\infty$.  
This is largely unsurprising as the prefactor on the deterministic term in Eq.~\eqref{eq:nondim combined} contains the product $\mu\nu$; taking $\mu\to 0$ while keeping this integral response factor relatively constant would necessitate $\nu\to\infty$.
Fig.~\ref{fig:contour fig}b shows a corresponding decrease in $\tau$, limiting towards pure diffusion.  
Returning to Eq.~\eqref{eq:nondim coupled PDE}, as $\mu\to 0$ both the diffusion and the source term go to zero, thus the concentration field would remain fixed in time.  
If this initial concentration field was constant, then the particle would have no gradient to respond to and therefore only undergo pure Brownian motion in this infinite memory regime, corresponding to $\tau=0$.
This suggests that rather than trying to start at finite $\mu$ and witness the effects of self-avoidant memory fade as $\mu\to\infty$, as this model was set up to do, future work should perhaps remove $\mu$ from the source term in Eq.~\eqref{eq:nondim coupled PDE} and start at $\mu=0$ to witness the effects of self-avoidant memory fade as $\mu$ increases away from zero. 

The limiting ballistic motion when taking both $\mu$ and $\nu$ to infinity is in contrast to the limiting Brownian motion behavior of Eq.~\eqref{eq:nondim combined} as $\mu\to \infty$  while keeping $\nu$ fixed.
By following contours of $\nu$ in Fig.~\ref{fig:contour fig}c, we see that the velocity first increases with $\mu$ and then decreases, approaching zero velocity as $\mu\to\infty$, which is consistent with the trend shown in Fig.~\ref{fig:paths}b.  
It is interesting to observe in Fig.~\ref{fig:contour fig}d that $\tau$ appears to be relatively static along the contours of $\nu$. 
Note that these values of $\tau$ were mainly computed at points to the left of the maximum velocity of the fixed $\nu$ contours.
Figure~\ref{fig:paths}b indicated that $\tau$ decreases with increasing $\mu$ and fixed $\nu$.  
When we also consider a decrease velocity, the trajectories can be assumed to approach Brownian motion.

In summary, increasing $\mu$ to decrease the effects of memory either results in increasing $\tau$ (by fixing $V$) and therefore creating straighter trajectories that do not display enhanced diffusion in the MSD over the timescale of the simulation, or in decreasing $V$ to zero (by fixing $\nu$) which results in a purely diffusive MSD.
The memory is responsible for both the ballistic motion measured by velocity $V$ and the effective rotational diffusion measured by orientational persistence time $\tau$, so naturally follows that these effects are both lost with increasing $\mu$.  
If the concentration field diffuses infinitely fast by taking $\mu \rightarrow \infty$ with $\nu$ fixed, we lose deterministic motion since the gradient of the concentration field is always zero at the particle's center and radially isotropic around the particle, thus the net force acting on the particle is always zero.  
The effective angular diffusion is lost when taking  $\mu \rightarrow \infty$ with $V$ fixed because this requires large $\nu$ such that the immediate deterministic forces overwhelm the noise and any past history, and so reduce the MSD to almost exclusively ballistic motion.
Thus, incorporation of self-avoidant memory is not simply an addendum to the active Brownian model that can be removed without consequence; by its complex interactions with the enhanced diffusion we see that it makes for a categorically unique model.

\section{\textbf{Conclusions}\label{sec:conclusions}}

We have analyzed the self-avoidant memory effects of a model coupling an active swimmer and an environmental chemical field. 
Like the experimental system it was inspired by, it can exhibit ABP-like behavior with the MSD having both a ballistic and a long-time enhanced diffusion regime.  
With an analytical formula for the velocity, $V$, we faithfully reproduced the ballistic regime.
The enhanced diffusion in our model is a result of both angular persistence and the self-avoidant memory, whereas ABP only includes orientational persistence.  
We found that numerically computing the 
 orientation decorrelation (or persistence) time, $\tau$,  enhanced diffusion predicted by ABP overestimates the enhanced diffusion in our model. 
Thus, our proposed model did not faithfully capture the dynamics of the experimental system at long time scales in the same way that ABP did. 
(Further investigation will be needed to determine if this difference is due to parameter values, modeling choices like using thermal noise and the diffusive scaling to the source term, or the absence of hydrodynamic effects.)
Instead, we discovered that the self-avoidant memory in our model led to transient self-trapping that suppressed the enhanced diffusion. 
This self-trapping  has, to date, been suggested to occur only in self-attracting experimental systems ~\cite{Liebchen_and_Lowen, Tsori} and computational models \cite{Lowen_2, Grima, Golestanian}.
Further investigation will be needed to determine if self-trapping is a unique feature of this model, or can occur in other self-avoidant systems.

Through these investigations, we kept the noise parameter $\epsilon$ fixed, while changing the gradient response parameter $\nu$ and the diffusion $\mu$ to find that both latter parameters control the implicit parameters $V$ and $\tau$. 
With only two control parameters, we were unable to independently tune each timescale of behavior: the velocity $V$, the memory timescale $\mu^{-1}$, and the angular persistence timescale $\tau$.  
Taking $\mu\to\infty$ to remove memory effects, we either arrived at simple Brownian motion by fixing $\nu$ or purely ballistic motion by fixing $V$ and allowing $\nu\to\infty$; the memory is responsible for both the ballistic motion and the effective rotational diffusion.
Taking $\mu\to 0$, we again arrive at simple Brownian motion, having removed all self-avoidant memory with our choice of scaling the source term in the concentration field by $\mu$.  
We thereby identified an intermediate regime of $\mu$ for which enhanced diffusion is present on a finite timescale, but at a lower magnitude than expected for ABP with equivalent angular persistence.  
This regime will be used in future work to study self-avoidant memory effects in many-particle simulations, investigating motility induced phase separation and associated dynamic pattern formation, which is commonly observed in active systems with particles that are repulsive to one another.

\begin{acknowledgments}
KAN thanks Eric Vanden-Eijnden for initial discussions concerning the model.
\end{acknowledgments}

\onecolumngrid
\appendix

\section{Solving the Diffusion Equation to Combine and Nondimensionalize the Coupled System\label{App:Solve}}

Consider the dimensional, 2D system given in Eq.~\eqref{eq:coupled}.
Taking the Fourier Transform of Eq.~\eqref{eq:coupled PDE} we arrive at the ODE 
\begin{equation}\label{eq:FT of PDE}
   \hat{c}_t +D|\mathbf{k}|^2\hat{c} =\frac{\alpha D R^2}{2\pi}\exp\left[-\frac{R^2}{2}\mathbf{\lvert k \rvert}^2+i(\mathbf{k}\cdot \mathbf{X}(t)) \right] .
\end{equation}

\noindent
We compute the integrating factor of Eq.~\eqref{eq:FT of PDE} which is  $I=e^{\int D \mathbf{\lvert k \rvert}^2 dt}=e^{Dt\mathbf{\lvert k \rvert}^2}$. From this Eq.~\eqref{eq:FT of PDE} can be rewritten as
\begin{equation}\label{eq:Int factor ODE}
     \frac{d}{dt}\left(  e^{ Dt|\mathbf{k}|^2}\hat{c}\right)=\frac{\alpha D R^2}{2\pi} e^{Dt|\mathbf{k}|^2}e^{\left[-\frac{R^2}{2}|\mathbf{k}|^2+i(\mathbf{k}\cdot \mathbf{X}(t)) \right] } .
\end{equation}

\noindent
Integrating both sides of Eq.~\eqref{eq:Int factor ODE} gives the solution to Eq.~\eqref{eq:FT of PDE}
\begin{equation}\label{eq:chat solution}
     \hat{c}= \frac{\alpha D R^2}{2\pi} \int_{0}^{t} e^{-D(t-s)|\mathbf{k}|^2}e^{\left[-\frac{R^2}{2}|\mathbf{k}|^2+i(\mathbf{k}\cdot \mathbf{X}(s)) \right] }ds .
\end{equation}

\noindent
Taking the inverse Fourier Transform of Eq.~\eqref{eq:chat solution} yields the solution, $c(\mathbf{x},t)$, to Eq.~\eqref{eq:coupled PDE}, which is

\begin{equation}\label{eq:c solution}
    c = \frac{\alpha D R^2}{2\pi}\int_{0}^{t}(R^2 +2D(t-s))^{-1}\cdot e^{-\frac{|\mathbf{x}-\mathbf{X}(s)|^2}{2(R^2 +2D(t-s))}}ds .
\end{equation}

\noindent
 We can incorporate the solution to Eq.~\eqref{eq:coupled PDE}, which is Eq.~\eqref{eq:c solution}, into Eq.~\eqref{eq:coupled SDE} by taking the gradient, $\nabla c$, which is
\begin{equation}
    \nabla c  = -\frac{\alpha D R^2}{2\pi}
   \int_{0}^{t}(R^2 +2D(t-s))^{-2}(\mathbf{x}-\mathbf{X}(s)) \cdot  e^{-\frac{|\mathbf{x}-\mathbf{X}(s)|^2}{2(R^2 +2D(t-s))}} ds .
\end{equation}

\noindent
The SDE path evolution Eq.~\eqref{eq:coupled SDE} then becomes
\begin{equation}\label{eq:combine SDE nonreduced}
    d\mathbf{X}=\frac{\alpha D \beta R }{(2\pi)^{2}} \left [ \int_{0}^{t}  (R^2 +2D(t-s))^{-2} \left (\int_{\mathbb{R}^2}(\mathbf{x}-\mathbf{X}(s))\cdot e^{-\frac{|\mathbf{x}-\mathbf{X}(s)|^2}{2(R^2 +2D(t-s))}-\frac{|\mathbf{x}-\mathbf{X}(t)|^2}{2R^2}}
    d \mathbf{x}\right )ds \right ] dt   + \sqrt{\sigma}d\mathbf{W} .
\end{equation}

\noindent
Evaluation of the spatial integral over $\mathbb{R}^2$ reduces Eq.~\eqref{eq:combine SDE nonreduced} to
\begin{equation}\label{eq:combine SDE intermediate}
   d\mathbf{X}=\frac{\alpha D \beta R^3 }{2^3 \pi} \left[\int_{0}^{t}  \left ( (\mathbf{X}(t)-\mathbf{X}(s))e^{-\frac{|\mathbf{X}(t)-\mathbf{X}(s)|^2}{4(R^2 +D(t-s))}}(R^2 +D(t-s))^{-2}\right)ds\right]dt   + \sqrt{\sigma}d\mathbf{W} .
\end{equation}

\noindent
By nondimensionalizing under the scalings $\mathbf{Y}=\frac{\mathbf{X}}{R}$, $\tau= \frac{t}{T}$,  and $\mathbf{B}=\frac{\mathbf{W}}{\sqrt{T}}$, Eq.~\eqref{eq:combine SDE intermediate} becomes

\begin{equation}\label{eq:combine SDE nondim unsimplified}
    R d\mathbf{Y}=\frac{\alpha D \beta R^3 }{2^3 \pi} \left[\int_{0}^{\tau}  \left ( e^{-\frac{|R\mathbf{Y}(\tau)-R\mathbf{Y}(\zeta)|^2}{4(R^2 +DT(\tau-\zeta))}}(R\mathbf{Y}(\tau)-R\mathbf{Y}(\zeta))(R^2 +DT(\tau-\zeta))^{-2}\right)Td\zeta\right]Td\tau   + \sqrt{\sigma T}d\mathbf{B} .
\end{equation}

\noindent
The SDE path evolution given by Eq.~\eqref{eq:combine SDE nondim unsimplified} then simplifies to
\begin{equation}
    d \mathbf{Y}=\frac{\alpha D \beta R^3 T^2}{ 2^3 \pi} \left[\int_{0}^{\tau}  \left ( e^{-\frac{|R\mathbf{Y}(\tau)-R\mathbf{Y}(\zeta)|^2}{4(R^2 +DT(\tau-\zeta))}}(\mathbf{Y}(\tau)-\mathbf{Y}(\zeta))(R^2 +DT(\tau-\zeta))^{-2}\right)d\zeta\right]d\tau   + \frac{\sqrt{\sigma T}}{R}d\mathbf{B} 
\end{equation}

\noindent
%where we also introduce the nondimensional dummy time variable $\zeta  = \frac{s}{T}.$ Since random variables are by definition dimensionless, a random variable $Z(t)$ is not a function of $t$; it is a draw from the distribution of $Z$ at time $t$. We want to then be able to draw from the same random variable at nondimensional time $\tau$, thus we map $\mathbf{X}(t)\rightarrow \mathbf{U}(\tau)$. Similarly, we want to evaluate the now-dimensionless integrand over its associated dimensionless time interval, so we set the upper limit of integration to $\tau$. 

\noindent
Incorporating the nondimensional parameters $D\rightarrow \mu=\frac{DT}{R^2}$,  $\alpha \rightarrow \phi = \frac{\alpha R^2}{2\pi}$, $\beta \rightarrow \nu =\frac{\beta T}{2\pi  R}$, and $\sigma \rightarrow \epsilon=\frac{\sigma T}{R^2}$ and exchanging $s$ for $\zeta$ and $t$ for $\tau$ for notational convenience we have the nondimensional SDE path evolution equation

\begin{equation}
d \mathbf{Y}=\frac{\pi}{2}\mu \nu \phi \left[\int_{0}^{t}  \left ( e^{-\frac{|\mathbf{Y}(t)-\mathbf{Y}(s)|^2}{4(1 +\mu(t-s))}}(\mathbf{Y}(t)-\mathbf{Y}(s))(1 +\mu(t-s))^{-2}\right)ds\right]dt   + \sqrt{\epsilon} d\mathbf{B}
\end{equation}
in agreement with Eq.~\eqref{eq:nondim combined}.

\bigskip
\noindent
\section{Computation of the Velocity Integral Formulation Using a Dirac Delta Function\label{App:dirac}}

To assess the case in which the particle is considered a point source, we substitute the mollified delta function, $\delta_R(\mathbf{x}-\mathbf{X}(t))= \frac{1}{2\pi R^2}e^{-\frac{|\mathbf{x}-\mathbf{X}(t)|^2}{2R^2}}$, in Eq.~\eqref{eq:nondim coupled} for a Dirac delta function,  
\begin{subequations}\label{eq:Dirac dimensional system}
\begin{equation}\label{eq:Dirac dimensional PDE}
    \frac{\partial c}{\partial t}= D \Delta c + \alpha D \delta^2(\mathbf{x}-\mathbf{X}(t))
\end{equation}
\begin{equation}\label{eq:Dirac dimensional SDE}
    d\mathbf{X}(t)=-\beta R \left(\int_{\Omega}\delta^2(\mathbf{x}-\mathbf{X}(t))\nabla c d\mathbf{x} \right)dt + \sqrt{\sigma}d\mathbf{W}.
\end{equation}
\end{subequations}
Here, $\delta^2(\mathbf{x}-\mathbf{X}(t))$ is a 2-dimensional Dirac delta function centered at $\mathbf{X}(t)$. The $R^2$ in the source term of the original PDE given by Eq.~\eqref{eq:coupled PDE} is no longer necessary. Accordingly, the units of $\alpha$ are $[\alpha]=c$ and the units of $\beta$ remain $[\beta]=\frac{L}{cT}.$
Nondimensionalizing Eq.~\eqref{eq:Dirac dimensional system} with the scalings  $\mathbf{y}=\frac{\mathbf{x}}{R}$, $\mathbf{Y}=\frac{\mathbf{X}}{R}$, $\tau= \frac{t}{T}$,
and $\mathbf{B}=\frac{\mathbf{W}}{\sqrt{T}}$
and where $\mu=\frac{DT}{R^2}$, $\phi = \frac{\alpha}{2 \pi}$, $\nu = \frac{\beta T}{R 2 \pi}$ and $\epsilon = \frac{ \sigma T}{R^2}$ we arrive at the new system
\begin{subequations}\label{eq:Dirac nondim system}
\begin{equation}\label{eq:Dirac nondim PDE}
    \frac{\partial c}{\partial t}= \mu \Delta c + 2 \pi \mu \phi \delta^2(\mathbf{y}-\mathbf{Y}(t))
\end{equation}
\begin{equation}\label{eq:Dirac nondim SDE}
    d \mathbf{Y}(t)=- 2\pi \nu \left( \int_{\Omega}\delta^2(\mathbf{y}-\mathbf{Y}(t)) \nabla c d\mathbf{y} \right)dt + \sqrt{\epsilon} d\mathbf{B}
\end{equation}
\end{subequations}
where $c$, $t$ and  $\Omega$  are  re-used  for  their  non-dimensional versions  for  convenience.

As in the case with the sized particle, we take the Fourier Transform of the Eq.~\eqref{eq:Dirac nondim PDE} to arrive at the ODE
\begin{equation}\label{eq:FT of Dirac PDE}
    \hat{c}_t+ \mu \mathbf{\lvert k \rvert}^2 \hat{c}= \mu \phi e^{i \mathbf{k}\cdot \mathbf{Y}(t)} .
\end{equation}
We compute the integrating factor of Eq.~\eqref{eq:FT of Dirac PDE} which is $I=e^{\int \mu \mathbf{\lvert k \rvert}^2 dt }=e^{\mu t \mathbf{\lvert k \rvert}^2}$. From this, Eq.~\eqref{eq:FT of Dirac PDE} can be rewritten as
\begin{equation}\label{eq:int factor ODE Dirac}
    \frac{d}{dt}\left( \hat{c}\cdot e^{\mu t \mathbf{\lvert k \rvert}^2}\right) = \mu \phi e^{i \mathbf{k}\cdot \mathbf{Y}(t)} \cdot e^{\mu t \mathbf{\lvert k \rvert}^2} .
\end{equation}
Integrating both sides of Eq.~\eqref{eq:int factor ODE Dirac} gives
\begin{equation}\label{eq:Dirac chat}
    \hat{c}=\mu \phi \int_{0}^{t}e^{-\mu(t-s)\mathbf{\lvert k \rvert}^2+ i\mathbf{k}\cdot\mathbf{Y}(s)}ds .
\end{equation}
We take the inverse Fourier Transform of Eq.~\eqref{eq:Dirac chat} to find the solution to Eq.~\eqref{eq:Dirac nondim PDE}, which is
\begin{equation}\label{eq:Dirac c solution}
    c=\mu \phi \int_{0}^{t}\left(2\mu(t-s) \right)^{-1}e^{-\frac{|\mathbf{y}-\mathbf{Y}(s)|^2}{4(\mu(t-s))}}ds .
\end{equation}

 We  incorporate the solution to Eq.~\eqref{eq:Dirac nondim PDE} into Eq.~\eqref{eq:Dirac nondim SDE} by computing the gradient $\nabla c$ of Eq.~\eqref{eq:Dirac c solution}, which is
\begin{equation}
    \nabla c= -\mu \phi \int_{0}^{t}\frac{(\mathbf{y}-\mathbf{Y}(s))}{(2\mu(t-s))^2 }\exp\left[{-\frac{|\mathbf{y}-\mathbf{Y}(s)|^2}{4(\mu(t-s))}}\right]ds .
\end{equation}
Eq.~\ref{eq:Dirac nondim SDE} then becomes
\begin{equation}\label{eq:Dirac nondim combined intermediate}
 d\mathbf{Y}(t)=  \nu \mu \phi\frac{\pi}{2 }\int_{0}^{t} (\mu(t-s))^{-2}\int_{\Omega} \delta^2(\mathbf{y}-\mathbf{Y}(t)) (\mathbf{y}-\mathbf{Y}(s))\exp\left[{-\frac{|\mathbf{y}-\mathbf{Y}(s)|^2}{4(\mu(t-s))}}\right] d\mathbf{y}ds dt +\sqrt{\epsilon} d\mathbf{B} .   
\end{equation}
Evaluation of the spatial integral over $\mathbb{R}^2$ reduces Eq.~\eqref{eq:Dirac nondim combined intermediate} to
\begin{equation}\label{eq:Dirac nondim final}
     d\mathbf{Y}(t) = \nu \mu \phi\frac{\pi}{2 }\int_{0}^{t} (\mu(t-s))^{-2} (\mathbf{Y}(t)-\mathbf{Y}(s))\exp\left[{-\frac{|\mathbf{Y}(t)-\mathbf{Y}(s)|^2}{4(\mu(t-s))}}\right] ds dt +\sqrt{\epsilon} d\mathbf{B} .
\end{equation}

Now, suppose that: $\mathbf{Y}(t)=\langle Vt, 0 \rangle$. This simplifies Eq.~\eqref{eq:Dirac nondim final} to

 \begin{equation}\label{eq:Dirac V int}
     \frac{d\mathbf{Y}(t)}{dt}=V= \nu \mu \phi\frac{\pi}{2 }\int_{-\infty}^{t} (\mu(t-s))^{-2}(Vt-Vs)\exp\left[{-\frac{|Vt-Vs)|^2}{4(\mu(t-s))}}\right]ds .
 \end{equation}
By making the change of variables given by $z= \mu(t-s)$ and $ ds= -\frac{1}{\mu}dz$, we see that Eq.~\eqref{eq:Dirac V int} is considerably reduced to
\begin{equation}
    V=\nu\mu \phi\frac{\pi}{2} \int_{0}^{\infty} z^{-1}\frac{V}{\mu} \exp\left[ -\left( \frac{V}{\mu}\right)^2 \frac{z}{4} \right]dz .
\end{equation}
After some further simplification we arrive at the following expression,
\begin{equation}\label{eq:Dirac V int final}
    1=\frac{\pi}{2}\frac{\nu}{\mu}\phi\int_{0}^{\infty}\frac{1}{z}\exp\left[ -\left( \frac{V}{\mu}\right)^2 \frac{z}{4} \right]\frac{1}{\mu}dz .
\end{equation}

\noindent
This integral on the right hand side is not pointwise convergent for finite $V$ and thus indicates that when we consider the particle to be a point source with the self-avoidant memory that we have defined, the particle does not swim.

%%%%%%%%%%%%%%%%%%%%%
\section{Computation of the Hover Height Integral Formulation\label{App:hover}}

To show that the presented model also reproduces the experimentally-observed hovering of the droplets above the bottom place, we set the second component of the position $\mathbf Y$ in the direction perpendicular to the bottom plate, and add a constant non-dimensional gravitational force $f_g$.  
We then seek a steady state solution of the form $\mathbf{Y} = (0,h)$ for non-dimensional hover height $h$ of the droplet's center with reflecting boundary condition for the concentration field at $\mathbf y = (x_1, 0)$ with $x_1\in\mathbb{R}$.  Using the model formulation in Eq.~\eqref{eq:nondim combined} we can account for this boundary condition using the standard trick of placing an image particle at $\mathbf Y^*=(x_1,-x_2)$.  The resulting equation for the position $\mathbf Y$ including the image particle $\mathbf Y^*$ and the gravitational force is
\begin{equation} 
\begin{split}
 d\mathbf{Y}  &=   \frac{\pi}{2}\mu \nu \phi \int_{0}^{t}\exp \left[ -\frac{|\mathbf{Y}(t)-\mathbf{Y}(s)|^2}{4(1+ \mu(t-s))}\right] (1+ \mu(t-s))^{-2} (\mathbf{Y}(t)-\mathbf{Y}(s))ds  dt \\
 & + \frac{\pi}{2}\mu \nu \phi \int_{0}^{t}\exp \left[ -\frac{|\mathbf{Y}(t)-\mathbf{Y}^*(s)|^2}{4(1+ \mu(t-s))}\right] (1+ \mu(t-s))^{-2} (\mathbf{Y}(t)-\mathbf{Y}^*(s))ds  dt - (0,f_g) + \sqrt{\epsilon} d\mathbf{B}.
 \end{split}
\end{equation}
Isolating the second component, and looking for solutions $\mathbf Y=(0,h)$ and $\mathbf Y^*=(0,-h)$ for all time, with no noise ($\epsilon = 0$) we arrive at
\begin{equation}
    f_g =  {\pi}\mu \nu \phi \int_{-\infty}^{t}\exp \left[ -\frac{h^2}{(1+ \mu(t-s))}\right] (1+ \mu(t-s))^{-2} h ds . 
\end{equation}
Under the change of variables $z=\mu(t-s)$, the above is equivalent to
\begin{equation}\label{eq:hover}
    f_g = \pi \nu \phi \int_0^\infty \exp \left[ -\frac{h^2}{1+ z}\right] \frac{h}{(1+ z)^{2}}  dz
\end{equation}
which is independent of the memory timescale $\mu^{-1}$ as one might intuitively expect.  

\begin{figure}
     \centering
      \includegraphics[scale=0.7]{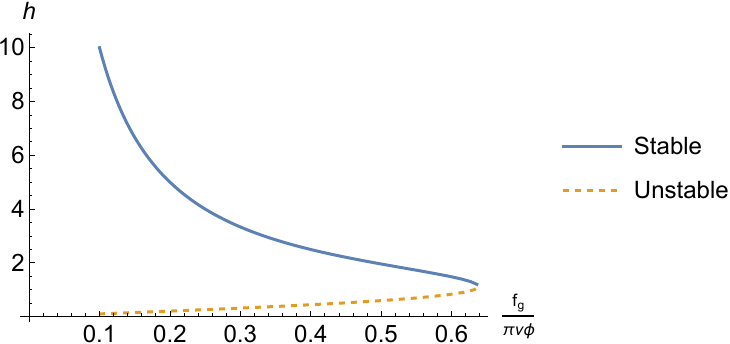}
     \caption{(color online) Solutions to Eq.~\eqref{eq:hover} for the hover-height $h$ of the droplet's center above the bottom plate as a function of $f_g/\pi\nu\phi$.  Beyond a critical value of $f_g/\pi\nu\phi$ (corresponding to $h\approx 1$) the droplets no longer hover but sit on the bottom plate.}
     \label{fig:hover}
 \end{figure}

Numerically-determined solutions to Eq.~\eqref{eq:hover} as a function of $f_g/\pi\nu\phi$ are shown in Fig.~\ref{fig:hover}.  Beyond a critical value of this parameter grouping, the droplets would no longer hover and rather fall to the bottom.  Note this occurs at about $h=1$ which is the non-dimensional radius $R$; the unstable solutions are within the fictitious boundary of the droplets.  A qualitative comparison to the experimental results of Fig.~3 in Ref.~\cite{Brujic2_autophoresis_synthetic_recent_1} reveals two similar trends.  First, increased SDS concentration yields a higher hover height.  In our model, this roughly corresponds to a stronger response to the concentration gradient, or the parameter $\nu$.  Increasing $\nu$ similarly increases the hover height.  Second, increased radius of the particles decreased the hover height.  In our model, this roughly corresponds to increasing the non-dimensional gravitational force $f_g$ which too decreases the hover height.

%%%%%%%%%%%%%%%%%%%%%
\bigskip
\noindent
\section{Computing the Small Time Asymptotics of the Active Brownian MSD\label{App:smallMSD}}

Recall the MSD given for the active Brownian particle (ABP) model with translational noise and rotational diffusion given in Eq.~\eqref{eq:ABP}:
\begin{equation}
       \mathbb{E}[\mathbf{X}(t)^2]=4V^2 \tau^2\left[2 \left(e^{-\frac{ t}{2\tau}}-1\right)+ \frac{ t}{\tau} \right] + 2 \epsilon t
\end{equation}
Starting from the MSD in Eq.~\eqref{eq:MSD ABP} for the ABP model, 
we rewrite the exponential as an infinite series to arrive at
\begin{equation}
    \mathbb{E}[\mathbf{X}(t)^2] = 4V^2 \tau^2\left[2 \left(\sum_{n=0}^{\infty}\frac{1}{n!}\left(-\frac{t}{2\tau}\right)^n  -1\right)+ \frac{ t}{\tau} \right] + 2 \epsilon t.
\end{equation}
This is asymptotic to 
\begin{equation}
  \mathbb{E}[\mathbf{X}(t)^2]  \approx 4V^2 \tau^2\left[2\left(\left( 1- \frac{t}{2\tau}+\frac{t^2}{8\tau^2}\right)-1\right)+ \frac{ t}{\tau} \right] + 2 \epsilon t
\end{equation}
as $t\to 0$ by just retaining a few leading order terms.

In the small time scale regime where $t^n \gg t^{n+1}$, we see that
\begin{equation}
     \mathbb{E}[\mathbf{X}(t)^2] \approx V^2t^2 + 2\epsilon t
\end{equation}
we obtain Eq.~\eqref{eq:MSD ABP small t}.
This expression is dominated by the diffusion-generated term $2\epsilon t$ at the smallest time scales (where $t\gg t^2$) and dominated by the directed motion term $V^2 t^2$ when $t^2$ becomes sufficiently larger than $t$.

Returning to Eq.~\eqref{eq:MSD ABP} in the large timescale regime where $t\gg\tau $, we see that
$$ e^{-\frac{t}{2\tau}}\rightarrow 0$$
and therefore
$$ \mathbb{E}[\mathbf{X}(t)^2] \approx (4V^2 \tau +2\epsilon)t $$
as given by Eq.~\eqref{eq:ABP MSD big t}.
This expression contains the amount of enhanced diffusion, $4V^2\tau^2$.

\section{Computing MSD and OCF from Position Time Series Generated by the Model\label{app:OCF}}

\begin{figure}
\includegraphics[width=0.95\textwidth,clip, trim=0 4cm 0 3cm]{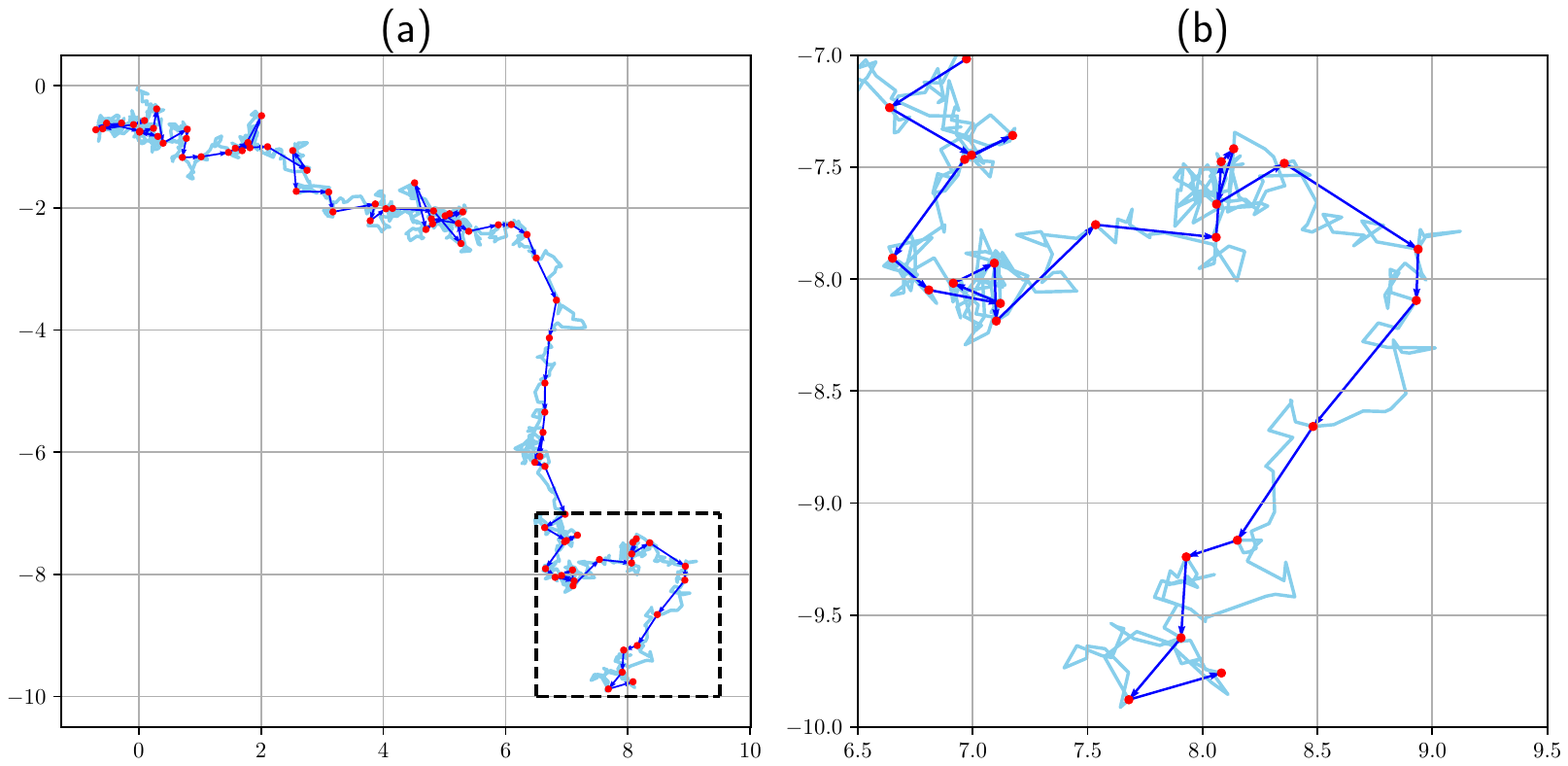}
\caption{Position time series of a sample trajectory with coarse grained lag times of 10 increments. Panel (a) includes entire trajectory and panel (b) is the inset identified with the dashed square.}
   \label{fig:vector fig}
\end{figure}

Absent a closed form expression for the mean square displacement of our model, we compute the empirical MSD from the position time series of length $N+1$ given by $\mathbf{X}(t)$: $\{ \mathbf{X}(0),\hdots,\mathbf{X}(N)\}$. 
To avoid introducing any correlations into the increment averages, we use non-overlapping increments. 
To achieve statistical accuracy, we then average over many simulated trajectories.
We denote the integer lag time as $\Delta L$ indicating the displacement traveled by the particle between observations $j$ and $j + \Delta L$ and given by $\mathbf{X}(j+ \Delta L)- \mathbf{X}(j)$. 
The total number of non-overlapping increments of length $\Delta L$ in a time series of length $N+1$ is $k=\lfloor \left(\frac{N+1}{\Delta L} \right) \rfloor$.
(In the event that the index lag length $\Delta L$ does not evenly divide the number of increments $N +1$, we remove the extra data from the beginning of the time.)
Thus, the empirical formula for the mean square displacement over the lag time $\Delta L$ of a single particle is given by
\begin{equation}
    \Delta L^2 = \frac{1}{k-1}\sum_{i=1}^{k}(\mathbf{X}(N-(i-1)\cdot\Delta L) - \mathbf{X}(N-i\cdot\Delta L))^2.
\end{equation}
 As shown in Fig.~\ref{fig:vector fig}, successive increases in $\Delta L$ result in a sampling process which coarse grains the position time series.

Using the same partitioning process described above and shown in Fig.~\ref{fig:vector fig} we can compute the non-overlapping displacements and find the cosine between consecutive pairs.
The resulting time average of these computed cosines gives the orientation correlation function, for which the formula is given in Eq.~\eqref{eq:OCF}.

\twocolumngrid

\end{document}